\documentclass[]{pasj01}

\usepackage{url}

\begin{document} 
\jyear{2018}

\title{In-flight Calibration of Hitomi Soft X-ray Spectrometer (3) Effective Area}

\author{
Masahiro \textsc{Tsujimoto}\altaffilmark{1},
\email{tsujimot@astro.isas.jaxa.jp}
Takashi \textsc{Okajima}\altaffilmark{2},
\email{takashi.okajima@nasa.gov}
Megan E. \textsc{Eckart}\altaffilmark{2}, 
\email{megan.e.eckart@nasa.gov}
Takayuki \textsc{Hayashi}\altaffilmark{2,3},
Akio \textsc{Hoshino}\altaffilmark{2},
Ryo \textsc{Iizuka}\altaffilmark{1},
Richard L. \textsc{Kelley}\altaffilmark{2}, 
Caroline A. \textsc{Kilbourne}\altaffilmark{2},
Maurice A. \textsc{Leutenegger}\altaffilmark{2,4}, 
Yoshitomo \textsc{Maeda}\altaffilmark{1},
Hideyuki \textsc{Mori}\altaffilmark{2},
Frederick S. \textsc{Porter}\altaffilmark{2},
Kosuke \textsc{Sato}\altaffilmark{5},
Toshiki \textsc{Sato}\altaffilmark{1,6},
Peter J. \textsc{Serlemitsos}\altaffilmark{2},
Andrew \textsc{Szymkowiak}\altaffilmark{7},
Tahir \textsc{Yaqoob}\altaffilmark{2,4}
}
\altaffiltext{1}{Japan Aerospace Exploration Agency, Institute of Space and Astronautical Science, Chuo-ku, Sagamihara, Kanagawa 252-5210, Japan}
\altaffiltext{2}{NASA's Goddard Space Flight Center, Greenbelt, MD 20771, USA}
\altaffiltext{3}{Department of Physics, Nagoya University, Chikusa-ku, Nagoya, Aichi 464-8602, Japan}
\altaffiltext{4}{University of Maryland, Baltimore County, Baltimore, MD 21250, USA}
\altaffiltext{5}{Department of Physics, Tokyo University of Science, Shinjuku-ku, Tokyo 162-8601, Japan}
\altaffiltext{6}{Department of Physics, Tokyo Metropolitan University, Hachioji, Tokyo 192-0397, Japan}
\altaffiltext{7}{Department of Physics, Yale University, New Haven, CT 06520-8120, USA}

\maketitle

\begin{abstract} 
 We present the result of the in-flight calibration of the effective area of the Soft
 X-ray Spectrometer (SXS) onboard the Hitomi X-ray satellite using an observation of the
 Crab nebula. We corrected for the artifacts when observing high count rate sources with
 the X-ray microcalorimeter. We then constructed a spectrum in the 0.5--20~keV band,
 which we modeled with a single power-law continuum attenuated by an interstellar
 extinction.  We evaluated the systematic uncertainty upon the spectral parameters by
 various calibration items. In the 2--12~keV band, the SXS result is consistent with the
 literature values in flux (2.20 $\pm$ 0.08 $\times$10$^{-8}$~erg~s$^{-1}$~cm$^{-2}$
 with a 1$\sigma$ statistical uncertainty) but is softer in the power-law index (2.19
 $\pm$ 0.11). The discrepancy is attributable to the systematic uncertainty of about
 $+$6/$-$7\% and $+$2/$-$5\% respectively for the flux and the power-law index. The
 softer spectrum is affected primarily by the systematic uncertainty of the Dewar gate
 valve transmission and the event screening.
\end{abstract}

\section{Introduction}\label{s1}
The Soft X-ray Spectrometer (SXS; \cite{kelley16}) onboard the Hitomi satellite
\citep{takahashi16} ceased its short life before it was fully commissioned due to the
loss of the spacecraft control. Still, the instrument proved its superb performance,
meeting and even partially exceeding the requirements and yielding scientific
results. We only have a limited data set during its 38-day operation in the orbit
\citep{tsujimoto16}, but it added new information that we were unable to obtain during
the decade-long preparation on the ground. To make the best use of it, and also to prepare
for future X-ray microcalorimeter missions, it is important to verify the
instrumental calibration using in-flight data. In this paper, we discuss the
effective area calibration of the SXS including its telescope (SXT-S; the soft X-ray
telescope for SXS; \cite{okajima16}).

The effective area is determined not only by the area of the X-ray mirrors but also by
the transmission of various filters, the quantum efficiency and event redistribution
of the detector, dead time due to the digital electronics, event screening, the
background, the telescope pointing, and the ray-trace modeling. The entire system
was not tested end-to-end on the ground. Thus, the total effective area calibration
could only be verified with in-flight data. The goal of this paper is to evaluate the
in-flight observation of a celestial source in light of the available component- and
subsystem-level calibration data (table~\ref{t02}).

\begin{table*}[!]
 \tbl{References relevant for the SXS and SXT-S effective area calibration}
 {
 \begin{tabular*}{\textwidth}{ll}
  \hline
  Item & References \\
  \hline
  Telescope pointing & \texttt{jaxa\_hitomi\_memo\_2016-001}\footnotemark[$a$] \\
  Mirror effective area & Ground (\cite{iizuka14,tsato16a}; Iizuka et al. JATIS, submitted) and in-flight \citep{okajima16}\\
  Mirror edges & Au L \citep{kikuchi16,maeda16} and M edges \citep{kurashima16}\\
  Point spread function & Ground \citep{tsato14,iizuka14,tsato16b,hayashi16} and in-flight \citep{maeda18}\\
  Ray-tracing & Yaqoob et al. (JATIS, submitted), \texttt{asth\_sxt\_caldb\_mirror}\footnotemark[$a$], \texttt{asth\_sxt\_caldb\_auxtrans}\footnotemark[$a$], \texttt{asth\_caldb\_telarea}\footnotemark[$a$]\\
  Gate valve & \citet{eckart16}, Eckart et al. (JATIS, submitted), \citet{hoshino17,yoshida17}, \texttt{asth\_sxs\_caldb\_gatevalve}\footnotemark[$a$]\\
  Aperture filters & \citet{eckart16,kilbourne16a}, Eckart et al. (JATIS, submitted), \texttt{asth\_sxs\_caldb\_blckfilt}\footnotemark[$a$] \\
  Quantum efficiency & \citet{eckart16}, Eckart et al. (JATIS, submitted), \texttt{asth\_sxs\_caldb\_quanteff}\footnotemark[$a$] \\
  Energy gain & \citet{eckart16,leutenegger16}, Eckart et al. (JATIS, submitted), \texttt{asth\_sxs\_caldb\_gainpix}\footnotemark[$a$] \\
  Line spread function & \citet{eckart16,leutenegger16}, \texttt{asth\_sxs\_caldb\_rmfparam}\footnotemark[$a$] \\
  Data processing & \citet{ishisaki16,angelini16}\\
  Time assignment & \citet{eckart16}, \texttt{asth\_sxs\_caldb\_coeftime}\\
  NXB & \citet{kilbourne16b,porter16,kilbourne18}\\
  \hline
 \end{tabular*}
 }
 \label{t02}
 \begin{tabnote}
  \footnotemark[$a$] Available at \url{https://heasarc.gsfc.nasa.gov/docs/hitomi/calib/hitomi_caldb_docs.html.}
 \end{tabnote}
\end{table*}

All ground calibration measurements were performed using the flight unit prior to the
launch \citep{eckart16}. The only exception is the Dewar gate valve, which was placed at
the top of the Dewar to keep the Dewar vacuum on the ground. It was planned to be opened
upon confirmation that the initial spacecraft outgassing of potential contaminants
ceased $\sim$2 months after the launch. The spacecraft was lost a few weeks before this
operation, thus all data were taken with the gate valve in the optical path. The gate
valve window was not fully calibrated on the ground. This choice was motivated by its
not being part of the nominal flight configuration as well as schedule
considerations. After the loss of the mission, we obtained transmission data on a
flight-spare unit manufactured using the same lot of Be as the flight unit
\citep{hoshino17,yoshida17}, and use these data in this paper.

For this paper, we use the Crab observation simply because we have no other options. It
was observed as a commissioning target of other instruments on the spacecraft and was
not intended as a calibration source for the SXS. It is a challenging source with a high
count rate, but at the same time, we can demonstrate how well we calibrate and model the
instrument with small statistical uncertainties.

\medskip

The Crab is one of the standard candles in the International Astronomical Consortium for
High Energy Calibration (IACHEC), and the results of other instruments are available for
comparison
\citep{willingale01,weisskopf04,kirsch05,kaastra09,weisskopf10,madsen15b}. The SXS has a
different set of features and systematic uncertainties compared to other X-ray
instruments participating in the cross-calibration campaign.

In fact, some features of the SXS are well-suited to observe the Crab nebula. First, it
is a non-dispersive spectrometer, so the spectral resolution is not compromised nor
complicated by the extended nature of the source. Second, it is a high-resolution
spectrometer with a line spread function (LSF) that is dominated by a narrow ($\sim$5 eV
FWHM) Gaussian core, and thus the spectrum in the low-energy channels is not very much
contaminated by the redistributed events from the high-energy channels. Third, it has a
very low non--X-ray background (NXB) at a level of $\lesssim$1 event per spectral
resolution (5~eV) per 100~ks exposure \citep{kilbourne18}. It also achieved a wide
high-energy band coverage far beyond the required limit of 12~keV. As a result, we
obtained a spectrum of the Crab up to $\sim$25~keV above the NXB. Fourth, the sampling
rate is much higher (12.5~kHz) than conventional X-ray CCD spectrometers as necessitated
for high-resolution microcalorimeter spectroscopy, photon pile-up is much less
serious. Fifth, the relative timing accuracy is better than 80~$\mu$s
\citep{leutenegger16}, which is sufficient to resolve the 34~ms pulse phases of the Crab
pulsar.

A few features of the SXS are not favorable for this source. The first is the coarse
spatial resolution of 1\farcm2 half power diameter \citep{okajima16}, which is
insufficient to spatially resolve the pulsar and the nebula components. We only assess
the spatially integrated spectrum within the 3\arcmin\ square field of view. Second, for
high count rate observations like this, we effectively lose exposure time for high
spectral resolution events owing to overlapping pulses and overloading of the CPUs in the
onboard digital processing unit. The third is the loss of effective area below
$\sim$2~keV due to the gate valve Be window. This made it difficult to constrain
the amount of interstellar extinction of order 
$\sim$10$^{21.5}$~cm$^{-2}$. These features are corrected or evaluated as sources of
systematic uncertainties in this paper.

\medskip

We start with a brief description of the data set (\S~\ref{s2}). We then evaluate the
systematic uncertainty by the individual causes (\S~\ref{s3}), and compare them with
other results (\S~\ref{s4}). The main results of this study are summarized in
\S~\ref{s5}. Throughout this paper, we used the HEASoft and CALDB releases on 2017 May
12 for the Hitomi collaboration, the pipeline data products version 03.01.006.007
\citep{angelini16}, and the \texttt{Xspec} spectral fitting package version 12.9.1
\citep{arnaud96}.

\section{Data}\label{s2}
\subsection{Observation}\label{s2-1}
\begin{figure}
 \begin{center}
  \FigureFile(80mm,80mm){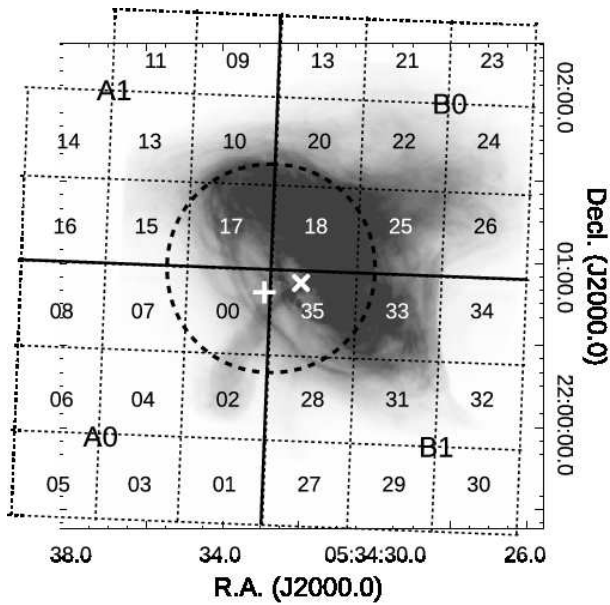}
 \end{center}
 \caption{Field of view of the four quadrants (A0, A1, B0, and B1) that comprises a
 6$\times$6 array (dotted squares with pixel numbers) superimposed on the Chandra ACIS
 image after correcting for the readout streaks \citep{mori04}. The top left pixel in
 the array is not read out to accommodate a dedicated calibration pixel (number 12) that
 does not receive celestial photons. A half power diameter circle of the SXS is
 shown around the array center with the dashed circle. The position of the optical
 axis and the pulsar \citep{lobanov11} are respectively shown with the plus and cross
 signs.}
 \label{f08}
\end{figure}

The observation was made on 2016 March 25 from 12:35 to 18:01 UT at the position of the
Crab pulsar (figure~\ref{f08}). The spacecraft revolved around the Earth 3.6 times and
experienced three Earth occultations and three South Atlantic Anomaly (SAA)
passages. The total telescope time was 21.5~ks, whereas the total on-source time was
9.7~ks. Because of the closed gate valve (\S~\ref{s3-4}), the effective area below
$\sim$2~keV was 0~cm$^{2}$. The actual incoming rate was 13\% of what we would expect if this
source were observed with the open gate valve.

We used cleaned events with an energy registration extended to 32~keV. We do not apply
the additional screening based on the time proximity of events among different pixels;
the false positive by this screening is too large for high count rate observations like
this (\S~\ref{s3-9}). Within the same pixel, events are graded depending on the relative
arrival time with the others. In brief, events are graded either into H (high-), M
(medium-), or L (low-resolution) for decreasing length to the closest events in
time. They are also graded either p (primary) or s (secondary) if it is the leading
pulse or not. When an event is not seriously overlapped by other events close in time,
the accuracy of the energy determination is high enough for high-resolution
spectroscopy. The Hp and Mp grades are recognized as such, which we call
``high-quality'' grades. The total number of cleaned events is 1.8$\times$10$^{6}$ with
an average count rate of 5.3~s$^{-1}$~pixel$^{-1}$. About 43\% of the events are of a
high-quality grade. We use events of all grades for a better statistics in this paper,
as we are mostly interested in the overall spectral shape.

\subsection{Time series}\label{s2-2}
\begin{figure*}[!]
 \begin{center}
  \includegraphics[width=1.0\textwidth, bb=0 0 842 595]{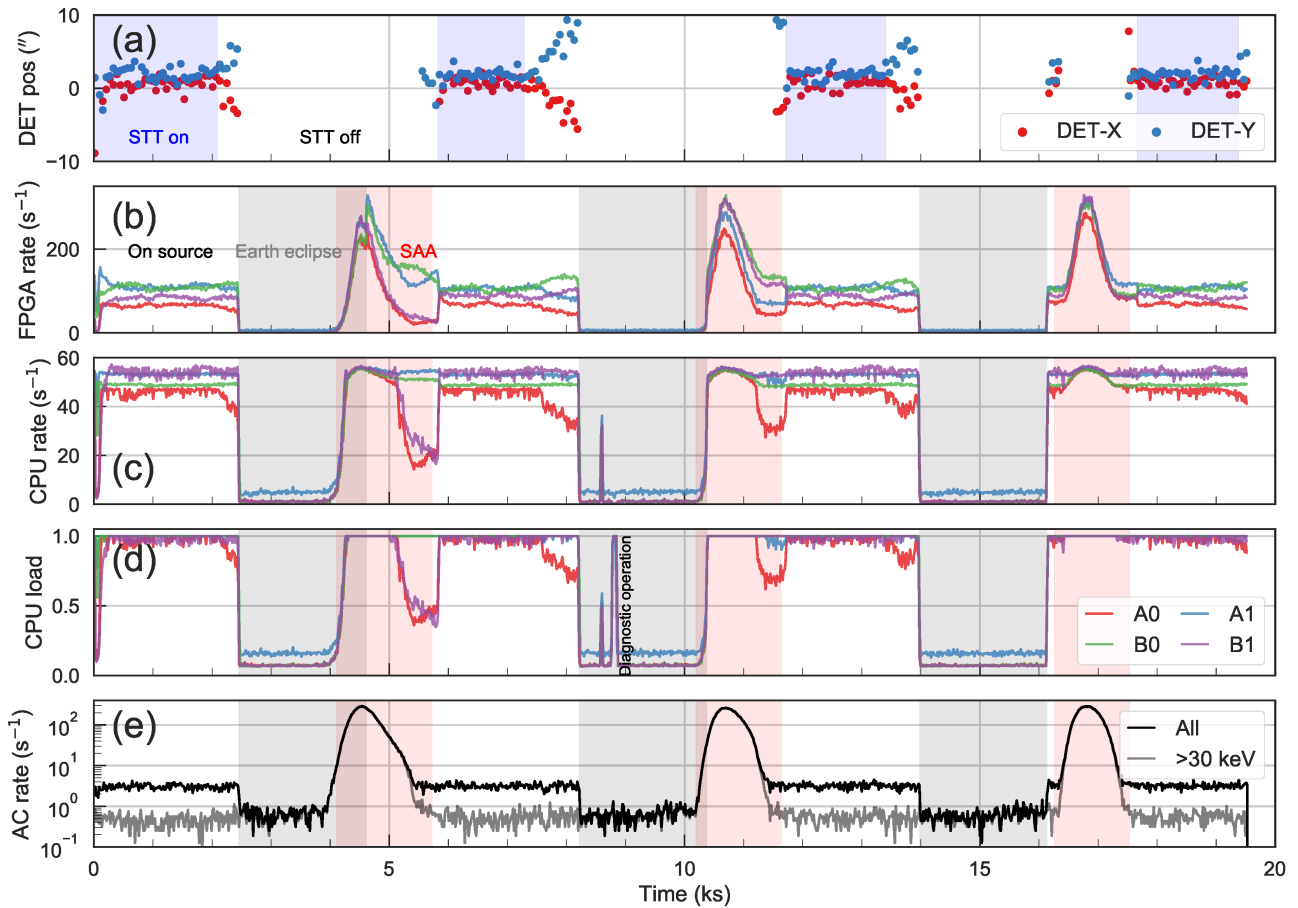}
 \end{center}
 \caption{Time series of the (a) median values of the detector coordinates of all events
 in a 30~s interval, (b) calorimeter event candidate rate detected by the FPGA, (c)
 calorimeter event rate processed by the CPU, (d) CPU load, and (e) anti-co event rate
 for all events (black) and those used for veto screening with energies above 30~keV
 (gray). The panels (b)--(d) are shown for the sum of all pixels in each of the four
 quadrants. Gray and red shades respectively indicate the duration of Earth eclipse and
 the SAA passages. The remainder is the on-source time. A part of it was operated with
 the STT, which is shown in the blue shade in panel (a). The spikes in (c) and (d) in
 8--9 ks are due to diagnostic operations. The start of the observation is 2016/03/25
 12:35:48 UT.}
 \label{f01}
\end{figure*}

The telescope pointing was measured and corrected using the star tracker (STT) and the
inertial reference unit. The STT achieves a better accuracy but was not in use
during the Earth occultations or SAA passages. The actual pointing fluctuation was
measured by calculating the median values of the detector coordinate of all X-ray events
in 30~s time slices during the on-source time (figure~\ref{f01}a). Some jumps in the
coordinate (e.g., 11.7 and 17.5~ks from the start) were seen when the STT service was
started and the pointing was quickly tuned. The degradation of the control is found when
STT was unavailable (e.g., 7.3--8.2~ks). When the STT was used, the pointing was
accurate to 0\farcs2 and was stable to 2\farcs9 at a 1$\sigma$ level. When the STT was
not used, the pointing was accurate to 3\farcs3 and was stable to 4\farcs0 at
1$\sigma$. These fluctuations are small enough in comparison to the detector pixel scale
of 30\arcsec, but it left a clear signature in the X-ray count rate from a source that
would otherwise be stable on a time scale of the observation.

The total throughput of the SXS for bright sources is determined by the CPU processing
speed of the onboard digital electronics called the Pulse Shape Processor (PSP;
\cite{ishisaki16}). The PSP consists of two identical units (PSP-A and PSP-B), and each
unit has one FPGA and two CPU boards. A total of 36 pixels is processed independently
with no priority for a particular pixel. All cross-channel processes including the
anti-coincidence (anti-co) veto are performed in the ground processing. A quarter of the
FPGA and CPU resources (called A0, A1, B0, and B1) handle nine pixels in a quadrant of
the array (figure~\ref{f08}). Pixel 12 is offset from the array and is illuminated by an
$^{55}$Fe calibration source with a constant rate of 3.9~s$^{-1}$, which is processed
with the A1 quadrant.

The FPGA board detects event candidates in the continuously input time series. Their
statistics are included in the house-keeping telemetry. The CPU board searches for
overlapping events upon the FPGA-detected event candidates, de-blends them, and derives
the arrival time and the energy by optimal filtering for each event, and assign flags
and grades to them. All these events are recorded in the event telemetry. The FPGA and
CPU rates are shown respectively in figure~\ref{f01} (b) and (c). The FPGA has a
sufficient buffer size for this observation, thus the rate is a proxy for the actual
incoming rate. On the other hand, the CPU speed was not fast enough and all four
CPUs were fully loaded during most of the Crab observation when the FPGA rate exceeded
$\sim$50~s$^{-1}$ per quadrant, as shown in figure~\ref{f01} (d).

At full load, the entire data buffer of up to 256 events stored by the FPGA is
occasionally discarded to catch up, which causes effective dead time. This is executed
pixel by pixel with no preference for a particular pixel, and lost times are
recorded. We call this ``pixel dead time'' and the fraction of its complement as ``live
time fraction'', hereafter. As a result, the rate of CPU-processed events
(figure~\ref{f01}c) is a fraction of the actual incoming rate. The anti-co events are
detected by the FPGA and are not processed by the CPU, thus have no loss of events
during the observation (figure~\ref{f01}e). High-energy events of the Crab were detected
by the anti-co, which is recognized by the elevated count rate during the on-source time
in comparison to the Earth eclipses for all events (black curve in
figure~\ref{f01}e). This discontinuity is not seen for anti-co events for those with
energies above 30~keV (gray curve), which are used for veto screening.

\subsection{Image}\label{s2-3}
\begin{figure*}[!]
 \begin{center}
  \includegraphics[width=0.45\textwidth, bb=0 0 360 360]{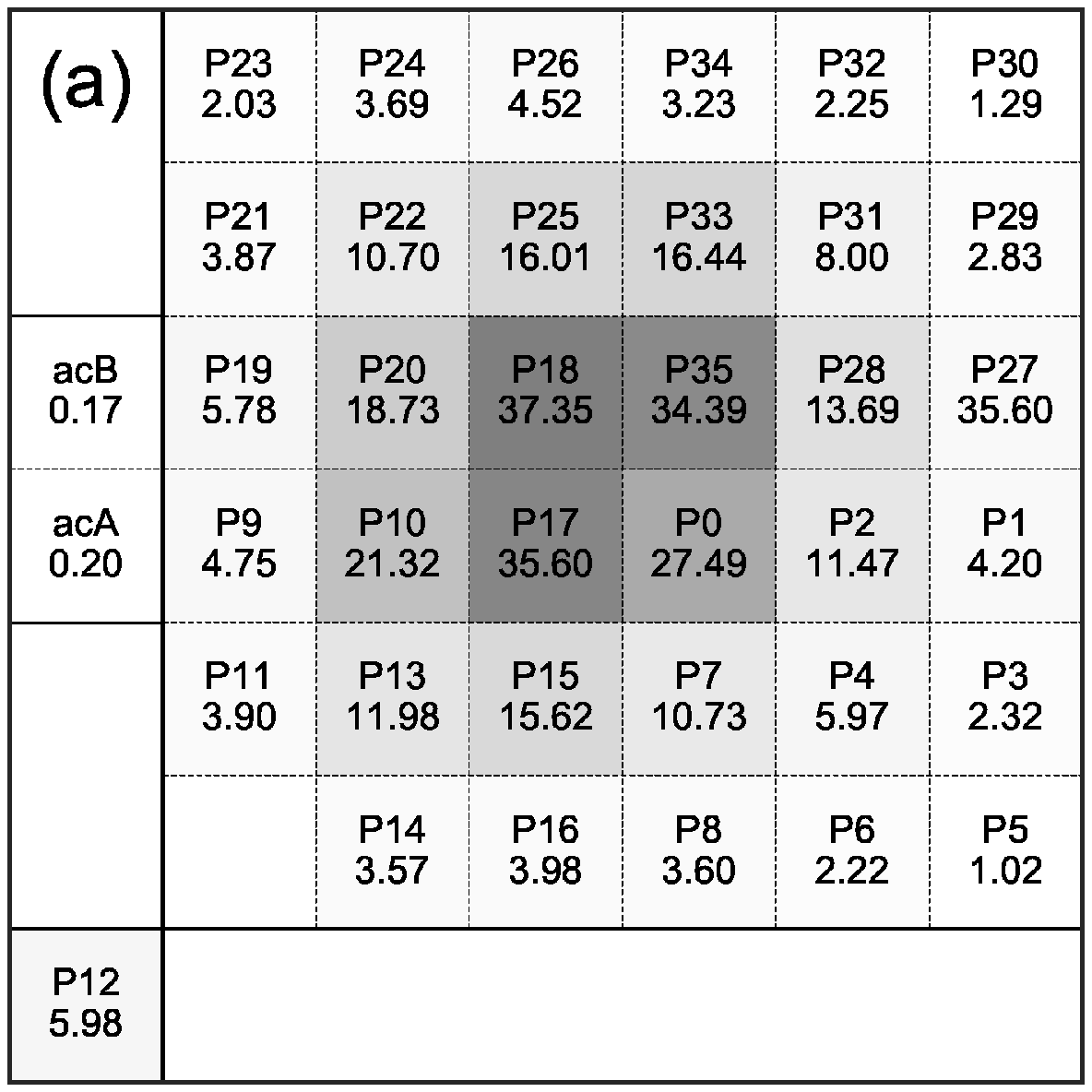}
  \includegraphics[width=0.45\textwidth, bb=0 0 360 360]{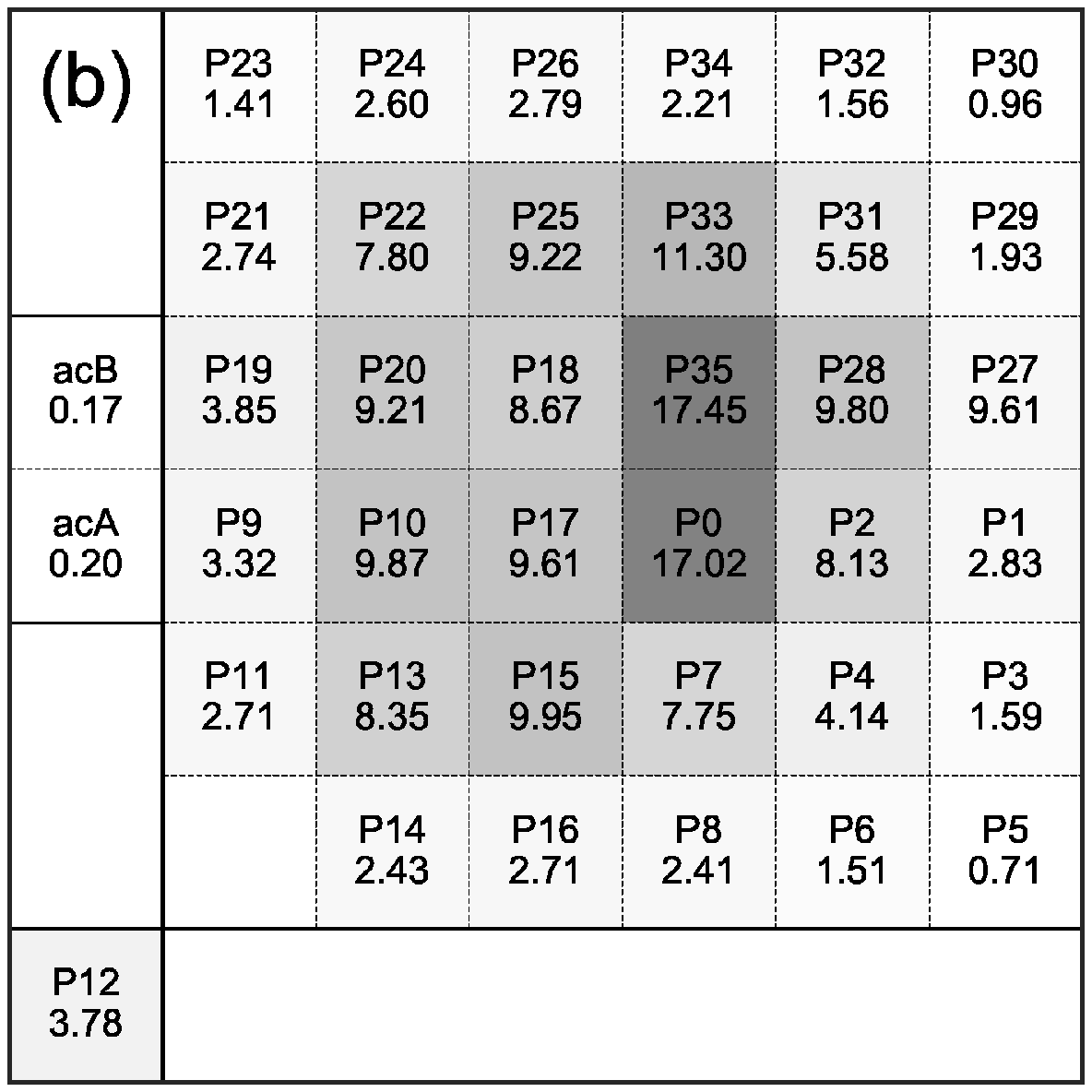}
  \includegraphics[width=0.45\textwidth, bb=0 0 360 360]{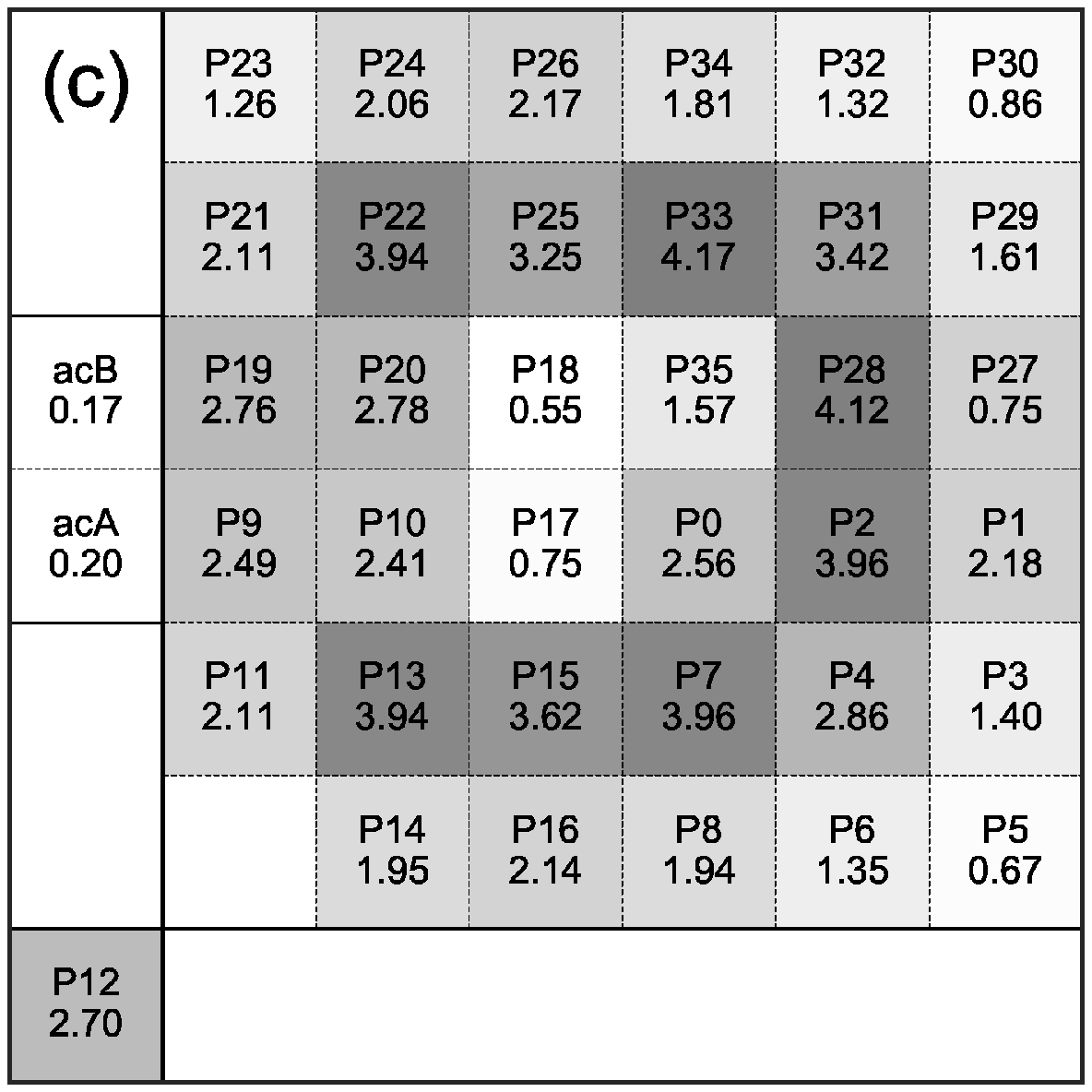}
  \includegraphics[width=0.45\textwidth, bb=0 0 360 360]{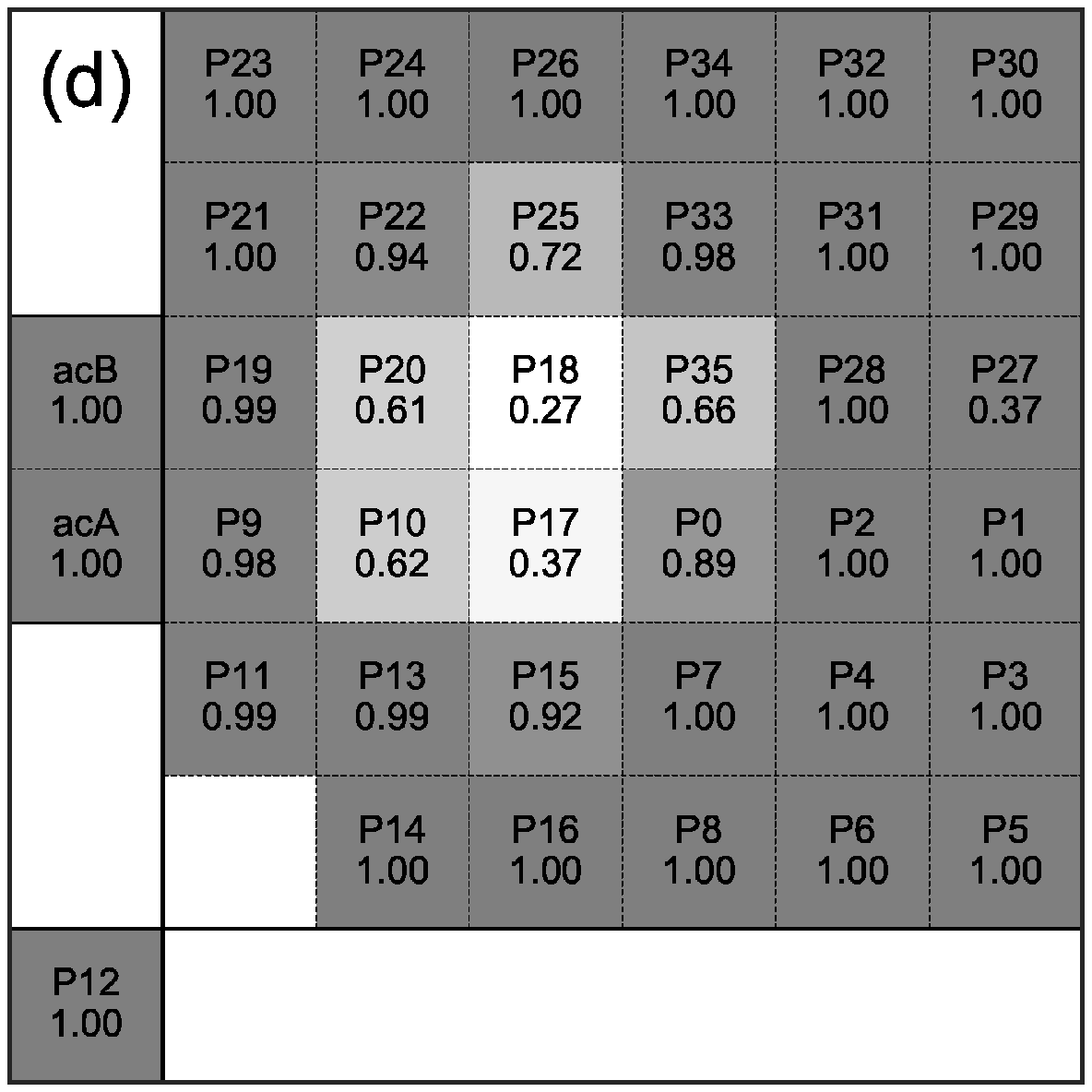}
  \includegraphics[width=0.45\textwidth, bb=0 0 360 360]{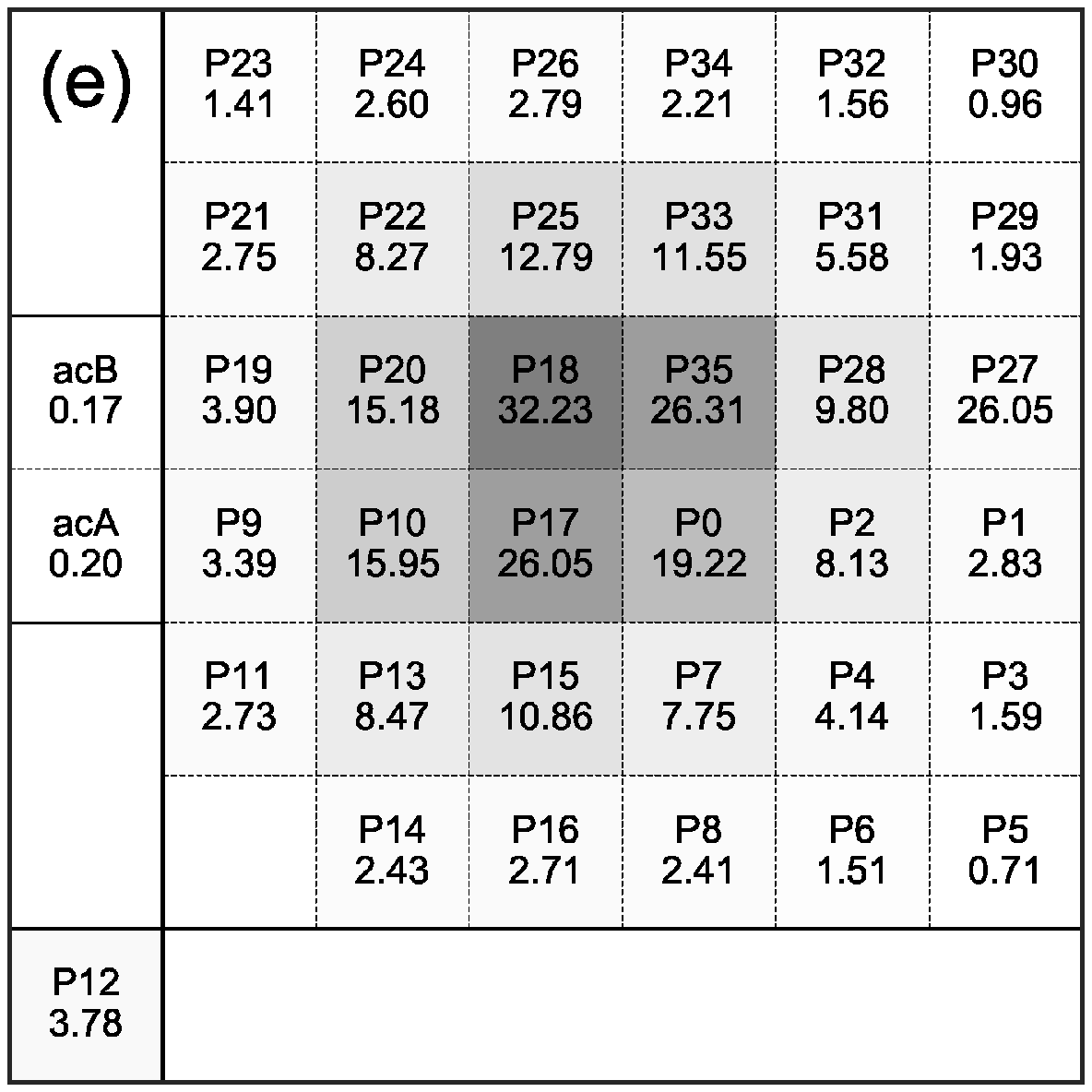}
  \includegraphics[width=0.45\textwidth, bb=0 0 360 360]{f02f.eps}
 \end{center}
 \caption{Count rate map in the unit of s$^{-1}$ of (a) event candidates detected by the
 FPGA, (b) events processed by the CPU not corrected for the pixel dead time, and (c)
 the events of a high-quality grade during on-source time, which is an average of the
 duration of 18--19~ks from the start of the observation (figure~\ref{f01}). (d) is the
 live time fraction of all on-source time. (e) processed event rate corrected for the
 pixel dead time by calculating (b) divided by (d). The maps are in the detector
 coordinate, as opposed to figure~\ref{f08} in the sky coordinate. The positions of the
 anti-co (acA and acB) and calibration (P12) channels are arbitrary.}
 \label{f02}
\end{figure*}

Figure~\ref{f02} shows the count rate map of (a) the event candidates detected by the
FPGA, (b) the events processed by the CPU, (c) the events of the high-quality grades,
and (d) the live time fraction during the on-source time. This illustrates various
artifacts when observing a bright source with the SXS. Because of the pixel dead time,
more illuminated pixels suffer a larger loss of events, thus the live time fraction is
lower at the center of the array (figure~\ref{f02}d). The map of events actually
processed by the CPU (figure~\ref{f02}b) should be corrected for this to derive the map
of actual rate incident on the detector (figure~\ref{f02}e). The event candidate rate by
FPGA (figure~\ref{f02}a) does not exactly match with figure~\ref{f02}e, as
figure~\ref{f02}a is a rate of candidates without inspecting the shapes of individual
pulses. When the incoming rate is too high, the rate of high-quality grades
decreases. When all these are combined, the map of the high-quality grade events has a
ring-like structure (figure~\ref{f02}c).

\subsection{Spectrum}\label{s2-4}
\begin{figure}
 \begin{center}
  \includegraphics[width=0.5\textwidth, bb=0 0 595 1263]{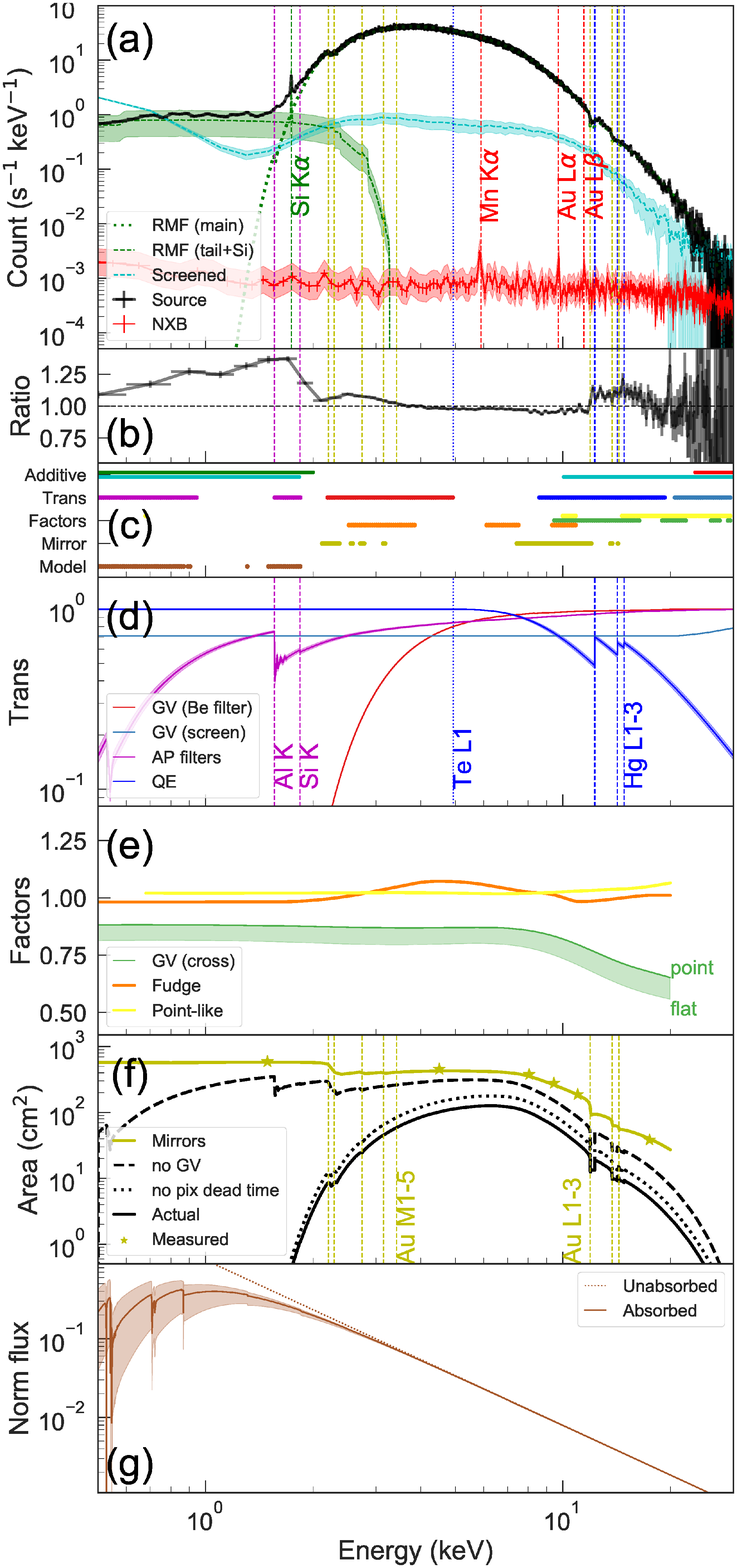}
 \end{center}
 \caption{(a) Source (black) spectrum and its decomposition into the main portion of the
 response (dominated by Gaussian core; dotted green) and the redistribution tail (dashed
 green), as well as the NXB spectrum (red) and the spectrum of the screened events
 (cyan), are shown. Instrumental features are labeled with the color corresponding to
 the relevant part of the spectrum. (b) Ratio of the background-subtracted spectrum and
 the canonical model (table~\ref{t03}) of the Crab nebula. (c) Energy band in which the
 individual causes are important. (d) Transmission of the gate valve Be filter and the
 support mesh structure, the aperture filters, and the quantum efficiency of the
 detector. (e) Correction function of the gate valve support cross structure, the SXT-S
 fudge function, and the extended nature of the source. (f) Effective area of the mirror
 assembly with the six measurements on the ground (yellow; Iizuka et al. JATIS,
 submitted) and the actual effective area (black solid) as well as that expected without
 the gate valve (dashed black) and without the pixel dead time (dotted black). (g)
 Spectral model. Both the unabsorbed and the absorbed power-law models are shown. The
 shaded range indicates the adopted uncertainty of the absorption.}
 \label{f03}
\end{figure}

Figure~\ref{f03} (a) shows the source and NXB spectra. The source spectrum was
integrated from the entire field of view and averaged over the pulse phases. The NXB
dominates the background of the SXS. The source has an excess signal against the NXB up
to $\sim$25~keV. As the effective area drops sharply below $\sim$2~keV, events below the
energy are mostly redistributed events from high-energy channels.

We compared the NXB-subtracted spectrum with a spectral model of the Crab nebula. The
Crab X-ray emission mainly consists of (i) the point-like, pulsed, and harder emission
from the pulsar and (ii) the extended, unpulsed, and softer emission from the
synchrotron nebula. However, the spectrum integrated over the space and time can be
simply described by a power-law of an index 2.1 without a break for the XMM-Newton
EPIC-pn (0.7--10~keV), RXTE PCA (3--60~keV), and NuSTAR (3--78~keV) spectra
\citep{weisskopf10,shaposhnikov12,madsen15}, so we used a single power-law continuum
attenuated by an interstellar extinction.

For the power-law component, we used the \texttt{pegpwrlw} model, in which the
unabsorbed flux in a given range is a free parameter. The flux is much better decoupled
with the power-law index than the conventional power-law model, in which the intensity
at 1~keV is a free parameter. For the extinction model, we used the \texttt{tbabs} model
version 2.3.2\footnote{See
\url{http://pulsar.sternwarte.uni-erlangen.de/wilms/research/tbabs/} for details.}
\citep{wilms00}. As we have no sensitivity below $\sim$2~keV due to the closed gate
valve, we cannot constrain the extinction column ($N_{\mathrm{H}}$). We
thus fixed this parameter to be 4.2$\times$10$^{21}$~cm$^{-2}$ with the oxygen abundance
decreased to 0.676 \citep{weisskopf04} relative to solar \citep{wilms00}.

We call this model the canonical model, in which the flux is the only free parameter. We
convolved the model with the instrumental response and fitted it to the data. The
best-fit parameters are shown in table~\ref{t03}, while the ratio of the data against
the model is shown in figure~\ref{f03} (b). The auxiliary response function (ARF) and
the redistribution matrix function (RMF) were generated following the data analysis
guide\footnote{The document is available at
\url{https://heasarc.gsfc.nasa.gov/docs/hitomi/analysis/}.}. In the calculation of ARF,
we considered the extended structure of the Crab nebula using the X-ray image taken with
the Chandra ACIS \citep{mori04} and the offset position of the optical axis from the
array center and the Crab pulsar position (figure~\ref{f08}). The redistribution into
the Si K fluorescence is not included in the RMF generator (\S~\ref{s3-7}), so we added
two Lorentzian models to represent the Si K$\alpha_{1}$ and K$\alpha_{2}$ lines.

\begin{table}[!ht]
 \tbl{Canonical model parameters in 0.5--24~keV.}
 {
 \begin{tabular}{lll}
  \hline
  Component  & Parameter & Value\footnotemark[$b$] \\
  \hline
  Power-law  & Index & 2.1 \\
             & Flux  & (4.947$\pm$0.037) $\times$ 10$^{-8}$ erg~s$^{-1}$~cm$^{-2}$\\
  Extinction & $N_{\mathrm{H}}$ & 4.2$\times$10$^{21}$~cm$^{-2}$\\
             & O abundance & 0.676 solar\\
             & Other abundance & 1.0 solar\\
  \hline
  Si K$\alpha$\footnotemark[$a$]  & K$\alpha_1$ energy & 1.73998 keV \\
                                  & K$\alpha_2$ energy & 1.73939 keV \\
                                  & Gain offset & --2.66$^{+0.33}_{-0.29}$~eV \\
                                  & K$\alpha_1$ intensity & 5.3$^{+0.7}_{-0.4}$ $\times$10$^{-2}$ cm$^{-2}$ s$^{-1}$\\
                                  & FWHM  & 1.3$^{+1.4}_{-0.75}$~eV \\
  \hline
  Fit goodness & Reduced $\chi^{2}$/dof & 1.12/23498\\
  \hline
 \end{tabular}
 }
 \label{t03}
 \begin{tabnote}
  \footnotemark[$a$] Two Lorentzian models were used to describe the natural line shape
  of Si K$\alpha_1$ and K$\alpha_2$. The two energies were allowed to shift collectively
  to adjust for the known SXS gain offset \citep{eckart16}. The K$\alpha_2$ intensity
  was fixed at half of that of K$\alpha_1$. These parameters were derived in the local
  fitting in 1.7--1.78~keV, then the best-fit values were used in the broadband
  fittings.\\ 
  \footnotemark[$b$] Errors indicate a 1$\sigma$ statistical
  uncertainty. Those without errors are fixed values.
 \end{tabnote}
\end{table}

\subsection{Pulse Phase}\label{s2-5}
\begin{figure}
 \begin{center}
  \includegraphics[width=0.5\textwidth, bb=0 0 842 595]{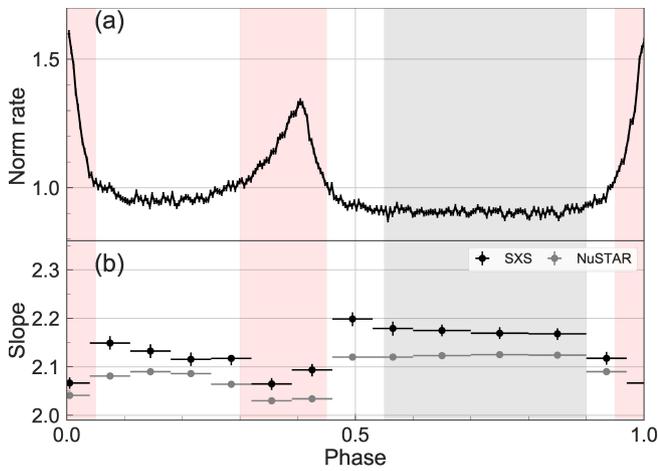}
 \end{center}
 \caption{(a) Folded light curve using events in the 2--20~keV band. Red and black shades
 define the on- and off-pulse phases, each of which occupies 35\% of a cycle. (b)
 Best-fit and 1$\sigma$ statistical uncertainty of the power-law index for phase-sliced
 spectroscopy of the SXS and NuSTAR \citep{madsen15}.}
 \label{f04}
\end{figure}

Figure~\ref{f04} (a) shows the light curve folded by the pulse period after a
barycentric correction for the position (RA, Dec) $=$ (\timeform{05h34m31.97232s},
\timeform{+22D00'52.069''}) in the equinox J2000.0. Events with all grades in the
2--20~keV band were used. The pulse period and its time derivative were derived as
33.7204626~ms and 4.198$\times$10$^{-13}$~s~s$^{-1}$ using the present data. The
ephemeris was determined based on simultaneous radio observations \citep{terada18}. The
SXS events are assigned times based on a clock synchronized to an onboard Global
Positioning System receiver. The accuracy of these time assignments is better than
20~$\mu$s \citep{leutenegger16}. The durations of the pixel dead times are longer than
5~s. Therefore, the 34~ms pulse profile is little distorted by the dead times.

\section{Analysis}\label{s3}
In this section, we evaluate the systematic uncertainties by various causes in the SXS
spectral fitting. Looking at the comparison between the data and the canonical model
(figure~\ref{f03}b), we find various deviations in the different energy bands, which
suggests that different causes of systematic uncertainties dominate in the different
bands. We divided the energy band into three: 2--4, 4--12, and 12--24~keV. In addition,
we evaluated in the energy band extended to the soft energies (0.5--12~keV), a part of
it limited by the gate valve (2--12~keV), and the entire band including the extended
parts on both soft and hard band (0.5--24~keV).

For each energy band, we fit the spectrum with the fiducial model, in which the photon
index was also treated as a free parameter in the canonical model. We evaluate the
effect of each individual source of systematic uncertainties
(\S~\ref{s3-1}--\S~\ref{s3-11}) in terms of the best-fit values of the free parameters
(photon index and the flux) in the fiducial model. Different values of $N_{\mathrm{H}}$
are considered as a source of systematic uncertainties (\S~\ref{s3-11}). The result is
summarized in each of the lines in table~\ref{t01}, while their effects are graphically
shown in the panels in figure~\ref{f03}.

\begin{table*}[ht]
 \tbl{Statistical and systematic uncertainties of the fiducial model fitting.}
 {
 \begin{tabular}{lcccccc|cccccc}
  \hline
  Parameter          & \multicolumn{6}{c|}{\rule{2cm}{.4pt}  Flux\footnotemark[$a$] \rule{2cm}{.4pt} } & \multicolumn{6}{c}{\rule{2cm}{.4pt}  Power-law index \rule{2cm}{.4pt} }\\
  Energy band (keV)  & 2--4 & 4--12 & 12--24 & 0.5--12 & 2--12 & 0.5--24 & 2--4 & 4--12 & 12--24 & 0.5--12 & 2--12 & 0.5--24 \\
  \hline
  Best-fit\footnotemark[$b$]    & 0.95 & 1.26 & 0.80 & 4.52 & 2.20 & 5.15 & 2.26 & 2.13 & 2.28 & 2.19 & 2.19 & 2.18  \\
  \hline
  Statistical (\%)\footnotemark[$c$] & $\pm$0.17 & $\pm$0.11 & $\pm$1.04 & $\pm$0.18 & $\pm$0.08 & $\pm$0.13 & $\pm$0.44 & $\pm$0.17 & $\pm$2.10 & $\pm$0.09 & $\pm$0.11 & $\pm$0.09 \\
  \hline
  Systematic (\%)\footnotemark[$d$] 
  & $^{+7.06}_{-13.1}$ & $^{+6.26}_{-4.23}$ & $^{+22.1}_{-9.22}$ & $^{+6.51}_{-13.2}$ & $^{+5.74}_{-6.55}$ & $^{+6.00}_{-10.5}$ 
  & $^{+6.74}_{-11.8}$ & $^{+4.58}_{-4.25}$ & $^{+8.27}_{-5.79}$ & $^{+1.94}_{-4.49}$ & $^{+2.09}_{-4.63}$ & $^{+2.52}_{-4.95}$\\
  \cline{2-13}
  ~~~Pointing\footnotemark[$e$] (\S~\ref{s3-1})
  & $-$0.49 & ($-$0.01) & ($-$0.14) & $-$0.53 & $-$0.16 & $-$0.39 & ($-$0.22) & ($+$0.01) & ($+$0.57) & $-$0.21 & $-$0.21 & $-$0.20 \\
  ~~~Image extent\footnotemark[$e$] (\S~\ref{s3-2})
  & $-$2.55 & $-$2.00 & $-$3.82 & $-$2.91 & $-$2.32 & $-$2.64 & ($+$0.19) & $-$0.46 & $+$2.37 & $-$0.34 & $-$0.35 & $-$0.29\\
  ~~~Mirrors\footnotemark[$e$] (\S~\ref{s3-3})
  & $+$2.13 & $+$4.11 & ($+$0.45) & $+$4.74 & $+$4.76 & $+$4.91 & $-$5.51 & $+$3.69 & $-$2.21 & ($+$0.01) & ($+$0.09) & $+$0.19 \\
  \cline{2-13}
  ~~~GV (Be filter)\footnotemark[$e$] (\S~\ref{s3-4})
  & $^{+3.85}_{-12.3}$ & $^{(+0.06)}_{-3.53}$ & $^{(+0.05)}_{-2.93}$ & $^{+3.54}_{-12.7}$ & $^{+0.80}_{-6.08}$ & $^{+2.50}_{-10.0}$ & $^{+6.09}_{-10.2}$ & $^{(+0.13)}_{-1.34}$ & $^{(+0.41)}_{(-0.11)}$ & $^{+1.46}_{-4.19}$ & $^{+1.31}_{-3.96}$ & $^{+1.36}_{-3.97}$\\
  ~~~GV (geom.)\footnotemark[$e$] (\S~\ref{s3-4})
  & --- & --- & --- & --- & --- & --- & $-$1.12 & $+$2.35 & $+$5.88 & $+$0.54 & $+$0.60 & $+$0.79 \\
  ~~~Aperture filters\footnotemark[$e$] (\S~\ref{s3-5})
  & $\pm$0.31 & $\pm$0.13 & ($\pm$0.01) & $\pm$0.34 & $\pm$0.20 & $\pm$0.29 & ($\pm$0.13) & ($\pm$0.06) & ($\pm$0.10) & ($\pm$0.08) & ($\pm$0.08) & ($\pm$0.08) \\
  ~~~Detector eff.\footnotemark[$e$]  (\S~\ref{s3-6})
  & ($\pm$0.01) & $\pm$1.01 & $\pm$3.32 & $\pm$0.81 & $\pm$0.35 & $\pm$0.43 & ($\pm$0.01) & $\pm$1.33 & ($\pm$0.62) & $\pm$0.66 & $\pm$0.68 & $\pm$0.71 \\
  \cline{2-13}
  ~~~LSF tail\footnotemark[$e$] (\S~\ref{s3-7})
  & $\pm$0.53 & $\pm$0.62 & $\pm$7.12 & $\pm$1.39 & $\pm$0.31 & $\pm$1.02 & $\pm$1.38 & ($\pm$0.10) & $\pm$5.20 & $\pm$0.78 & $\pm$0.48 & $\pm$0.91 \\
  ~~~Energy gain\footnotemark[$e$] (\S~\ref{s3-8})
  & --- & --- & $+$1.30 & --- & --- & --- & --- & --- & ($-$0.90) & --- & --- & ---\\
  \cline{2-13}
  ~~~Screening\footnotemark[$e$] (\S~\ref{s3-9}) 
  & $+$3.24 & $+$4.56 & $+$20.6 & $+$1.54 & $+$3.02 & $+$1.74 & $+$2.09 & $-$3.77 & ($-$0.15) & $-$1.06 & $-$1.38 & $-$1.65 \\
  ~~~NXB\footnotemark[$e$] (\S~\ref{s3-10})
  & ($\pm$0.01) & ($\pm$0.00) & ($\pm$0.37) & ($\pm$0.01) & ($\pm$0.00) & ($\pm$0.00) & ($\pm$0.01) & ($\pm$0.01) & ($\pm$0.58) & ($\pm$0.01) & ($\pm$0.00) & ($\pm$0.01) \\
  \cline{2-13}
  ~~~Spec model\footnotemark[$e$] (\S~\ref{s3-11})
  & $\pm$1.45 & $\pm$0.17 & ($\pm$0.01) & $\pm$1.50 & $\pm$0.53 & $\pm$1.13 & $\pm$1.40 & $\pm$0.17 & ($\pm$0.11) & $\pm$0.54 & $\pm$0.52 & $\pm$0.52 \\
  \hline
  Goodness of fit & \multicolumn{6}{c|}{\rule{2cm}{.4pt} Degree of freedom\footnotemark[$f$] \rule{2cm}{.4pt}} & \multicolumn{6}{c}{\rule{2cm}{.4pt} Reduced $\chi^{2}$ \rule{2cm}{.4pt}}\\
                  & 1997 & 7997 & 11997 & 11497 & 9997 & 23497 & 1.10--1.23 & 1.02--1.05 & 0.94--0.94 & 1.08--1.57 & 1.05--1.17 & 1.02--1.31\\
  \hline
 \end{tabular}
 }
 \label{t01}
 \begin{tabnote}
  \footnotemark[$a$] Absorption-corrected flux in the relevant energy band in the unit of 10$^{-8}$ erg~s$^{-1}$~cm$^{-2}$.\\
  \footnotemark[$b$] Best-fit values of the fiducial model.
  \footnotemark[$c$] The percentage of the 1 $\sigma$ statistical uncertainty with
  respect to the best-fit value of the fiducial model.\\
  \footnotemark[$d$] Quadrature sum of all the systematic uncertainty terms.\\
  \footnotemark[$e$] For each cause of the systematic uncertainty, the best-fit values
  are evaluated. (Evaluated value -- fiducial model value)/(fiducial model value) is shown in
  percentage. The values less than the statistical uncertainties indicate that the
  difference has no significance, which are shown in the parentheses.\\
  \footnotemark[$f$] Goodness of fit by the degree of freedom and the ranges of the reduced $\chi^2$ values.
 \end{tabnote}
\end{table*}

\subsection{Telescope pointing}\label{s3-1}
When the telescope optical axis is pointed off of a point-like source, a larger fraction
of photons land outside of the field of view, and thus the effective area
decreases. This is further complicated by the distribution of the pixel dead time
(figure~\ref{f02}d). Based on the telescope pointing accuracy (figure~\ref{f01}a), we
evaluated this effect by placing the optical axis position at offset places when
calculating the effective area.

\subsection{Image extent}\label{s3-2}
The effective area is calculated based on the Chandra image (figure~\ref{f08}). In order
to assess the systematic uncertainty by this assumption, we generated an ARF assuming
that the Crab nebula is a point-like source. The ratio of the effective area curve
assuming a point-like source with respect to that assuming the Chandra image
(figure~\ref{f08}) is shown in figure~\ref{f03} (e).

\subsection{X-ray mirror}\label{s3-3}
The effective area of the X-ray mirror assembly was measured on the ground at six
energies of the continuum up to $\sim$20~keV (\cite{tsato16a}; Iizuka et al. JATIS,
submitted) and additional detailed measurements of the Au L \citep{kikuchi16,maeda16}
and M \citep{kurashima16} edges. The Au L3 to L1 edges at 11.919, 13.734, and 14.353~keV
and the M5 to M1 edges at 2.206, 2.291, 2.743, 3.148, and 3.425~keV are the most
prominent features in the effective area curve of the Au-coated mirrors
(figure~\ref{f03}f), which are also recognized in the actual spectrum
(figure~\ref{f03}a).

The mirrors are characterized by various quantities such as the reflectivity, surface
roughness, degrees of misalignment, which are used by the ray-tracing simulator program
to calculate the mirror effective area. Because of the limited quantity of ground
calibration measurements and a large statistical uncertainty in the ray-tracing
calculation at small effective areas, no reliable effective area curve can be obtained
above $\sim$20~keV (Yaqoob et al. JATIS, submitted). Below $\sim$20~keV, discrepancies
between the ground-based effective area measurements and those from the ray-tracing
simulator up to $\sim$10\% are known. The discrepancy is much larger than the
statistical uncertainty in the ray-tracing output below $\sim$10 keV, but they are
comparable at $\sim$20 keV. The difference is interpolated to make a smooth correction
curve called the ``SXT-S fudge'' factor (figure~\ref{f03}e). We assessed the level of
systematic uncertainty by comparing the effective area curve with and without the fudge
factor.

\subsection{Gate valve}\label{s3-4}
The gate valve has three components that affect the effective area: (1) a Be filter that
allows some X-ray transmission, (2) a stainless steel protective screen, and (3) a cross
structure made of a thick Al for mechanical support \citep{eckart16}.

(1) The Be filter has a thickness of $\sim$262~$\mu$m and a density of 1.85~g~cm$^{-3}$
with a small amount of Mn, Fe, and Ni \citep{hoshino17,yoshida17}. The transmission
curve of the filter is characterized by the photo-electric absorption by Be and other
species, as well as the Bragg diffraction losses by Be. We made measurements of the
flight spare window at ground synchrotron facilities after the launch, which were
limited to energies above 2.5~keV. Among them, the measurements and the modeling below
3.8~keV remain highly uncertain. (2) The stainless screen covers 29\% of the geometrical
area and it reduces the flux by the fraction independent of the energy, except for
energies above $\sim$20~keV where it becomes partially transmissive. (3) The thick Al
support cross, which is opaque to X-rays, is aligned with the mirror quadrant gaps but
has a geometry such that it still blocks a significant fraction ($\sim$23.3\%) of
incident X-rays. The transmission curves of the Be filter and the stainless screen are
shown in figure~\ref{f03} (d).

There are two major sources of uncertainty. One is the Be filter transmission, which
dominates the total effective area curve below $\sim$5~keV (figure~\ref{f02}f). We
changed the transmission within the measurement errors and evaluated its effect. The
other is the geometrical effect by the support cross that partially intervenes the
X-rays far above the focal plane. As the higher energy X-rays are more preferentially
reflected by the inner part of the nested foils with smaller incident angles, they are
more likely to be blocked by the gate valve. The correction of this energy-dependent
factor was implemented in the ARF generator assuming that the source
is point-like (Yaqoob et al. JATIS, submitted). In theory, it depends on the source
distribution. We assessed its systematic uncertainty by comparing the correction curve
in two extreme cases that (i) the source is point-like and (ii) the source is uniform
across the SXS field of view. The curves for the two cases are shown in figure~\ref{f03}
(e). The ARF is generated to match with the assumed spatial distribution, thus we can
only compare the power-law index by this method.

\subsection{Aperture filters}\label{s3-5}
The SXS has a total of five filters in the X-ray light path. All five are made of an
Aluminized polyimide thin film with thicknesses of approximately 150--200 nm
($\sim$50--100~nm Al and $\sim$90--110~nm polyimide). They are anchored at different
temperature stages from 50~mK at the sensor thermal sink to $\sim$300~K at the Dewar
main shell. The outer three filters are supported by Si meshes. The combined
transmission is shown in figure~\ref{f03} (d) with the prominent Al and small Si K edge
features.

The combined filter transmission is calibrated much better than the required 5\% at 0.5,
1.5, and 6 keV \citep{eckart16}. We modified the thickness by $\pm$1\% for the entire
bandpass and estimated its effect.

\subsection{Detector efficiency}\label{s3-6}
The detector of the SXS consists of an array of 6$\times$6 pixels made by HgTe of
$\sim$10.5 $\pm$ 0.1~$\mu$m thickness. The quantum efficiency is 100\% below the energy
around the Te L1 edge at 4.949~keV, and monotonically decreases as the energy increases
except for the Hg L3 to L1 edges at 12.284, 14.209, and 14.839~keV (figure~\ref{f03}d). The
detector filling fraction is $\sim$90\%.

The efficiency is calibrated better than the required 5\% at 6~keV \citep{eckart16}. We
modified the absorber thickness by $\pm$5\% for the entire bandpass from the best-fit
value and estimated its effect.

\subsection{Extended LSF}\label{s3-7}
The detector response to monochromatic X-rays consists of (i) the main Gaussian peak
of $\sim$5~eV for high-quality grades, (ii) the exponential low-energy tail of an
$e$-folding scale of 12~eV, (iii) the escape peaks by Hg and Te, (iv) the electron loss
continuum, and (v) the X-ray fluorescence of Si \citep{leutenegger16,eckart16}. For (i)
and (ii), events are redistributed to the energies localized to the incident energy
(hereafter, the ``main'' components). For (iii) and (iv), redistributed events are spread
over a wide range of energies below the incident energy due to the energy loss either by
escaping fluorescent X-rays or electrons (the ``tail'' components). For (v), the
redistributed events are only significant at K$\alpha$ (1.74~keV).

In figure~\ref{f03} (a), we decomposed the observed spectrum into the main, tail, and Si
fluorescence components. The tail component is larger than the main component below
$\sim$1.8~keV since the gate valve Be window blocks low-energy celestial photons from
reaching the detector. The tail components are modeled using the fraction of the
electron loss continuum and 51 escape peaks, which amounts to a few percent of the main
peak in total. A detailed modeling is underway, but we currently have $\approx$50\%
uncertainty in these fractions \citep{leutenegger16}. We thus changed the fractions
collectively by $\pm$50\% by keeping the unity of the redistribution and evaluated its
effect in the fitting and assessed the corresponding uncertainty.

\subsection{Energy gain}\label{s3-8}
The energy gain scale is non-linear for the SXS. The gain curve for each pixel was
derived by fitting ground calibration data (4.5--13.5~keV) to a fourth order polynomial,
providing an accuracy of $\pm$ 2~eV across that bandpass \citep{eckart16}. These gain
curves were adjusted for flight operating conditions using limited on-orbit data
(primarily the $^{55}$Fe data on the calibration pixel and a single exposure of the main
array to calibration photons from the filter wheel $^{55}$Fe source;
\cite{leutenegger16}). Above the bandpass, the deviation from the model increases, which
alters the spectral shape of the model. It is estimated that the current gain is
under-estimated by 45--100~eV at 20~keV (Eckart et al. JATIS submitted). Assuming that a
gain shift of --100~eV at 20~keV, this changes the power-law index and the flux at
12--24~keV by $-$0.9\% and 1.3\% respectively.

\subsection{Event screening}\label{s3-9}
Events are screened using various flags in the pipeline processing. In the present
observation, 3.8\% of all recorded events were removed by this screening. The spectrum of the
screened events is shown in figure~\ref{f03} (a). There are two major types of flags;
one is based on the pulse shape of events and the other is based on the time
coincidence with other events. All screenings are intended to remove only spurious
events, but some X-ray events are removed by false positives. This effectively reduces
the exposure time and changes the spectral shape. Some false positives can be
calculated. For example, the anti-co veto screening reduces the exposure time by
$\sim$0.05\% for removing events arriving within $\pm$0.5~ms of an anti-co event above
30~keV at an average rate of 0.54~s$^{-1}$ during the Crab observation. Others are
more difficult to quantify.

\begin{figure}
 \begin{center}
  \includegraphics[width=0.5\textwidth, bb=0 0 595 842]{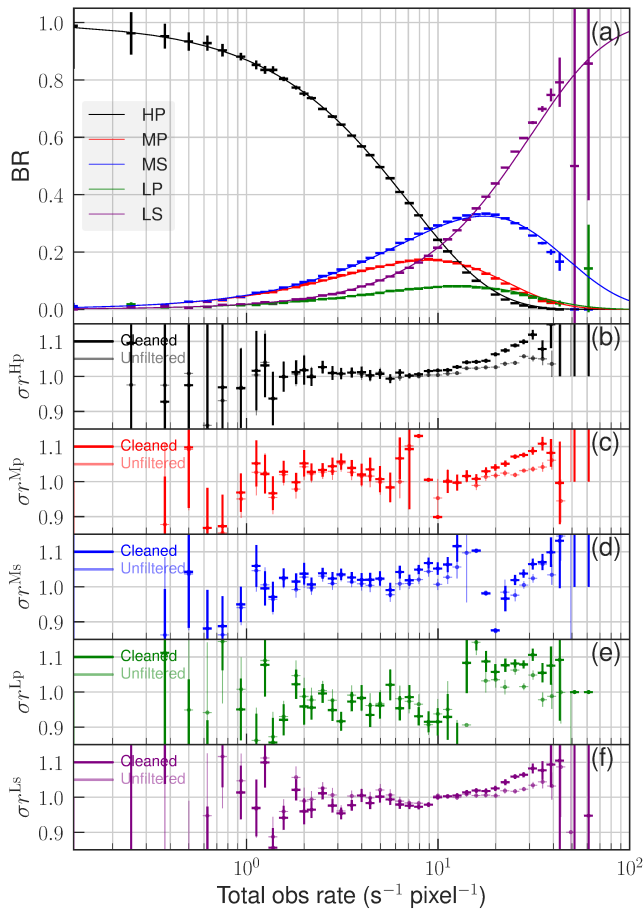}
 \end{center}
 \caption{(a) Observed branching ratio as a function of the observed total rate for the
 five grades (crosses) and the theoretical predictions assuming that the arrival time is
 random (solid curves). (b--f) For each grade, the ratio of the observed and theoretical
 total count rate is shown by using screened (dark) and unscreened (light color) events.}
 \label{f05}
\end{figure}

One way to assess the degree of unintended loss of events is to compare the observed and
theoretical incoming rate. The theoretical rate here is defined as an incoming rate that
explains the observed grade branching ratio based on the Poisson statistics. Assume that
the grade $g$ has a theoretical branching ratio of $P^{(g)}(r)$ at a total incoming rate
of $r$ (figure~\ref{f05}a). We used the cleaned events of the entire Crab observation to
derive the observed count rate of each grade $r^{(g)}_{\mathrm{obs}}$ at every 8~s
binning and grouped them by the total observed count rate defined by $r_{\mathrm{obs}} =
\Sigma_{g} r^{(g)}_{\mathrm{obs}}$. The binning is sufficiently large in comparison to
the pulse period, so that the arrival times can be considered to follow the Poisson
statistics. Here, we corrected the rates for the intensity-dependent pixel dead times.
By using all pixels, a wide range of $r^{(g)}_{\mathrm{obs}}$ can be covered. For each
group of an $r_{\mathrm{obs}}$ value, the total theoretical count rate
$r^{(g)}_{\mathrm{th}}$ was derived so that it satisfies $P^{(g)}(r^{(g)}_{\mathrm{th}})
= r^{(g)}_{\mathrm{obs}} / r_{\mathrm{obs}}$ for each grade $g$. The deviation is
defined as $\sigma r^{(g)} = r^{(g)}_{\mathrm{th}}/r^{(g)}_{\mathrm{obs}}$
(figure~\ref{f05}b--f).

The $\sigma r^{(g)}$ value is consistent with 1.0 at count rates lower than
10~s$^{-1}$~pixel$^{-1}$, indicating that the loss by event screening is negligible. It
increases above 1.0 significantly at a higher count rate up to $\sim$10\%, suggesting
that some X-ray counts are lost. At least a half of them can be explained by
screening. Indeed, if we perform the same exercise using the unscreened events, the
discrepancy decreases by about a half as shown in light colors in figure~\ref{f05}
(b)--(f). If two events of the same pixel are too close in time, they are not even
distinguished as two events. This effect is not large, however, as we see no trend of
grade migration from Ls or Lp to Hp.

X-ray events with higher energies are more prone to inappropriate flagging as they have
a longer pulse shape. Because of this, the spectrum can be distorted by unintended
screening. We evaluated this effect by comparing the fitting results of the spectra made
with unscreened or screened events.

\subsection{NXB}\label{s3-10}
The NXB spectrum was constructed by integrating the data during Earth pointing and
weighting by the history of the Earth magnetic cut-off rigidity. The spectrum is flat
over the entire energy band (figure~\ref{f03}a) with some features of Mn K$\alpha$ by
scattered X-rays from the $^{55}$Fe calibration source and Au L$\alpha$ and L$\beta$ by
fluorescence X-rays primarily from the Au layer of the sensor array frame \citep{kilbourne16b}.

We assumed that the spectrum above 30~keV is dominated by the NXB. The systematic
uncertainty of the NXB was evaluated by rescaling the NXB spectrum within the 1 $\sigma$
Poisson fluctuation of the counts above 30~keV. The NXB spectra were generated for the
minimum and maximum rescaling factors (0.56 and 1.74. respectively), which were
subtracted for the fitting to evaluate their systematic uncertainties. As the NXB is
almost negligible, there are no significant systematic uncertainties except for the high
energy band.

\subsection{Spectral model}\label{s3-11}
Because the SXS data has little constraining power for the amount of interstellar
extinction due to its low effective area below
$\sim$2~keV, we applied the extinction model that best describes the Chandra LETG
spectrum \citep{weisskopf04}. However, the LETG result is not exactly applicable to the
SXS spectrum primarily because the event extraction regions and the PSF are not the
same. The extinction column is known to vary across the nebula \citep{mori04}.

We evaluated how our fitting result is affected by changing the extinction column values
over (3.26--5.64) $\times$ 10$^{21}$~cm$^{-2}$ (figure~\ref{f03}g), which are two extreme
values in a cross-calibration study \citep{kirsch05}.

\section{Discussion}\label{s4}
\subsection{Major causes for the residuals}\label{s4-1}
We now assessed the effects of the systematic uncertainty by various conceivable causes
(table~\ref{t01}). Different causes leave a prominent effect in different energy
bands. We attempt to identify the primary causes for these residuals below.

First, for each one of the causes, we identified the energy band in which the cause can
change the spectral shape without changing the shape in other bands. We grouped
the causes into two: the ``additive'' causes and the ``multiplicative'' causes. The
former consists of the RMF tail (\S~\ref{s3-7}), the event screening (\S~\ref{s3-9}),
and the NXB (\S~\ref{s3-10}), and the latter consists of the others. For the additive
causes, the energy band was selected in which the additive spectrum exceeds 10\% of the
main component of the source spectrum (figure~\ref{f03}a). For the multiplicative
causes, the transmission, efficiency, or the correction factor (figure~\ref{f03}d, e, f,
and g) as a function of energy $f(E)$ is used to calculate $|df(E)/dE|$ at
logarithmically-equally spaced energies at 0.5--30~keV. Then, a 20\% of the energies
were selected from the top among those sorted with a decreasing order of
$|df(E)/dE|$. The energy bands are shown in figure~\ref{f03} (c).

The residual below $\sim$2~keV is attributable to the LSF tail (\S~\ref{s3-7}) or the
event screening (\S~\ref{s3-9}), which is larger than the other additive cause and the
main component of the spectrum (figure~\ref{f03}a). If we attribute the residual
entirely to the LSF tail, a $\sim$20\% increased value of the tail fraction gives the
smallest residual. The residual at $\sim$2--4~keV is attributable to the gate valve Be
filter (\S~\ref{s3-4}) or the mirror effective area around the Au M edges
(\S~\ref{s3-3}). The residual at $\sim$8--16~keV is attributable to the mirror effective
area (\S~\ref{s3-3}), the event screening (\S~\ref{s3-9}), the geometrical effect of the
gate valve (\S~\ref{s3-4}), or the detector efficiency (\S~\ref{s3-6}). They are hard to
be decoupled, but their fractional contribution can be assessed in the 4--12~keV fitting
result in table~\ref{t01}, in which the mirror effective area and the event screening
are the largest. It is illustrated by the change in the residual ratio from $<1$ to $>$1
(figure~\ref{f03}b) at the Au L3 edge attributable to the mirrors, but not at the Hg L3
edge attributable to the detector efficiency.  Together with the Au M features, the Au L
features of the mirror effective area is not easy to calibrate and model. A typical
systematic uncertainty around theses edges is estimated to be $\approx$10\%
\citep{kurashima16}.

Given the limited amount of in-flight data, it is difficult to disentangle these causes
further. For the other objects, multiplicative causes can be corrected by applying the
Crab ratio, which is defined as the ratio between the SXS spectrum and the canonical
model presented here\footnote{We provide the Crab ratio correction factor file, which is
available upon request.}.

Above $\sim$20~keV, the SXS is not very well calibrated, especially because of the lack
of measurement and hence modeling of the mirrors due primarily to the low mirror
effective area at high energies. Such high-energy bands are far beyond our
requirement. Although we have a clear signal above the NXB above 20~keV, the data cannot
be used for scientific purposes.

\subsection{Comparison with IACHEC results}\label{s4-2}
\begin{figure}
 \begin{center}
  \includegraphics[width=0.5\textwidth, bb=0 0 842 595]{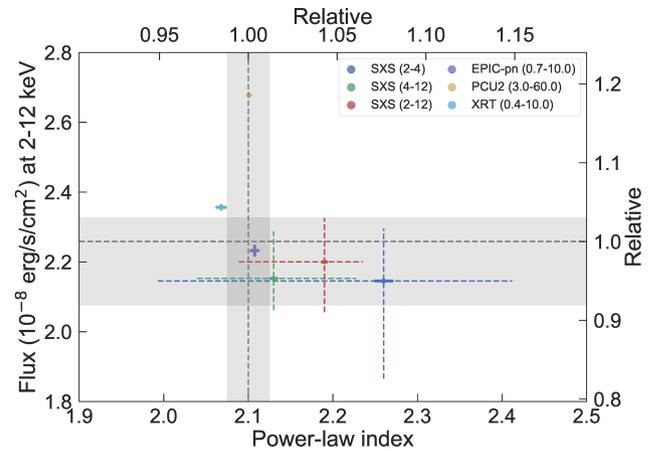}
 \end{center}
 \caption{Comparison between the SXS and the others \citep{weisskopf10,kirsch05} for the
 flux and the power-law index. The data are obtained by fitting in different energy
 bands, which is shown in the parentheses in the legend in the unit of keV. The flux
 values are converted to that in the 2--12~keV band, which is corrected for the
 absorption for this plot. For SXS, thick solid error bars are for the statistical
 errors, while thin dashed bars are for the quadrature sum of the systematic errors
 (table~\ref{t01}). Dashed gray lines indicate the values for the IACHEC model, whereas
 the shaded gray indicates their typical range of the intrinsic variation.}
 \label{f09}
\end{figure}

We compare our results with some preceding work using different instruments. In
figure~\ref{f09}, we show the best-fit values of the power-law index and the 2--12~keV
flux of the SXS (table~\ref{t01}) as well as those of
XMM-Newton EPIC-pn, RXTE PCU2 \citep{weisskopf10}, and Swift XRT \citep{kirsch05}. The
statistical errors are shown for the SXS and the other data, while the quadrature sum of
all the systematic errors (table~\ref{t01}) are shown for the SXS data.

The IACHEC model, in which all the parameters are fixed to those in \citet{weisskopf10},
is shown in gray dashed lines. The Crab spectrum is known to be variable both in the
flux and the power-law index \citep{wilson-hodge11}. We derived the typical variation of
the flux using the monitoring data with the Monitor of All-sky X-ray Image (MAXI;
\cite{matsuoka09}) and Gamma-ray Burst Monitor (GBM; \cite{meegan09}) onboard the Fermi
observatory, and that of the power-law index using RXTE PCA \citep{shaposhnikov12}. The
intrinsic variation is indicated with gray shades.

For the flux, the best-fit value in the fiducial model of the SXS is consistent with the
IACHEC model. On the other hand, that of the power-law index for the SXS is softer than
the IACHEC model and also with the other data. This is not due to the small extraction
region of the SXS. The softer index is consistently seen in all phase-resolved
spectroscopy (figure~\ref{f04}b), in which the contrast of the pulsar and the nebular
component changes. Also, when the extraction region is smaller, we would expect that the
index becomes harder \citep{madsen15} as the fraction of the harder pulsar component
increases. The statistical uncertainty is negligible. Therefore, it is likely that the
discrepancy is due to the systematic uncertainty of the SXS. In fact, the systematic
uncertainty of the power-law index, which is primarily due to the gate valve and event
screening calibrations, is large enough to explain most of the discrepancy between the
SXS result and the IACHEC model.

\section{Summary}\label{s5}
We presented the result of the in-flight calibration of the effective area of the SXS
using the Crab observations. The 0.5--20~keV spectrum was described by a single
power-law continuum attenuated by the interstellar extinction. We evaluated the level of
systematic uncertainty associated with various calibration items. A quadrature sum of
these uncertainties amounts to about $+$6/$-$7\% and $+$2/$-$5\% respectively for the
flux and the power-law index in the 2--12~keV band. The best-fit value of the flux with
the SXS is consistent with the others, whereas that of the power-law index is
softer. However, the discrepancy is mostly accountable within the systematic uncertainty
of the SXS. The primary causes of the softer spectrum of the SXS than the others are the
calibration of the gate valve transmission and the event screening.

\begin{ack}
 The authors gratefully acknowledge the many other scientists, engineers, technicians,
 and students whose hard work contributed to the capabilities of the SXS. We also
 appreciate Yuki Yoshida and Shunji Kitamoto at Rikkyo University for providing the
 spare Be filter transmission data. This work was supported by JSPS KAKENHI Grant
 Numbers JP15H03642, JP16K05309, JP25105516, JP23540280, and the Grant-in-Aid for
 Scientific Research on Innovative Areas Nuclear matter in neutron stars investigated by
 experiments and astronomical observations. T.\,S. is supported by the Research Fellow
 of JSPS for Young Scientists.
\end{ack}

\bibliographystyle{aa} 
\bibliography{ms} 

\end{document}